\documentclass[review]{elsarticle}

\usepackage{helvet}
\usepackage{courier}
\usepackage{amsmath,bm}
\usepackage{amssymb}
\usepackage{caption}
\usepackage{graphicx}
\usepackage{subfig}
\usepackage{multirow}
\usepackage{makecell} 
\usepackage{diagbox} 
\usepackage{cases}
\usepackage{enumerate}
\usepackage{color}
\usepackage[linesnumbered,ruled,vlined]{algorithm2e}

\newtheorem{proof}{Proof}

\journal{Journal of \LaTeX\ Templates}

\begin{document}

	\begin{frontmatter}
		
		\title{Latent Evolution Model for Change Point Detection in Time-varying Networks}

		\author[address1]{Yongshun Gong}
		
		\author[address1]{Xue Dong}
		
		\author[address2]{Jian Zhang}
		
		\author[address1]{Meng Chen \corref{mycorrespondingauthor}}
		\cortext[mycorrespondingauthor]{The corresponding author.}

		\address[address1]{School of Software, Shandong University, Jinan, Shandong, China}
		\address[address2]{University of Technology Sydney, Australian}

		\begin{abstract}
			Graph-based change point detection (CPD) play an irreplaceable role in discovering anomalous graphs in the time-varying network. While several techniques have been proposed to detect change points by identifying whether there is a significant difference between the target network and successive previous ones, they neglect the natural evolution of the network. In practice, real-world graphs such as social networks, traffic networks, and rating networks are constantly evolving over time. Considering this problem, we treat the problem as a prediction task and propose a novel CPD method for dynamic graphs via a latent evolution model. Our method focuses on learning the low-dimensional representations of networks and capturing the evolving patterns of these learned latent representations simultaneously. After having the evolving patterns, a prediction of the target network can be achieved. Then, we can detect the change points by comparing the prediction and the actual network by leveraging a trade-off strategy, which balances the importance between the prediction network and the normal graph pattern extracted from previous networks. Intensive experiments conducted on both synthetic and real-world datasets show the effectiveness and superiority of our model.
		\end{abstract}
		
		
	\end{frontmatter}

	\section{Introduction}\label{sec:introduction}
	
\begin{figure}[t]
		\centering
		\includegraphics[width=0.8\linewidth]{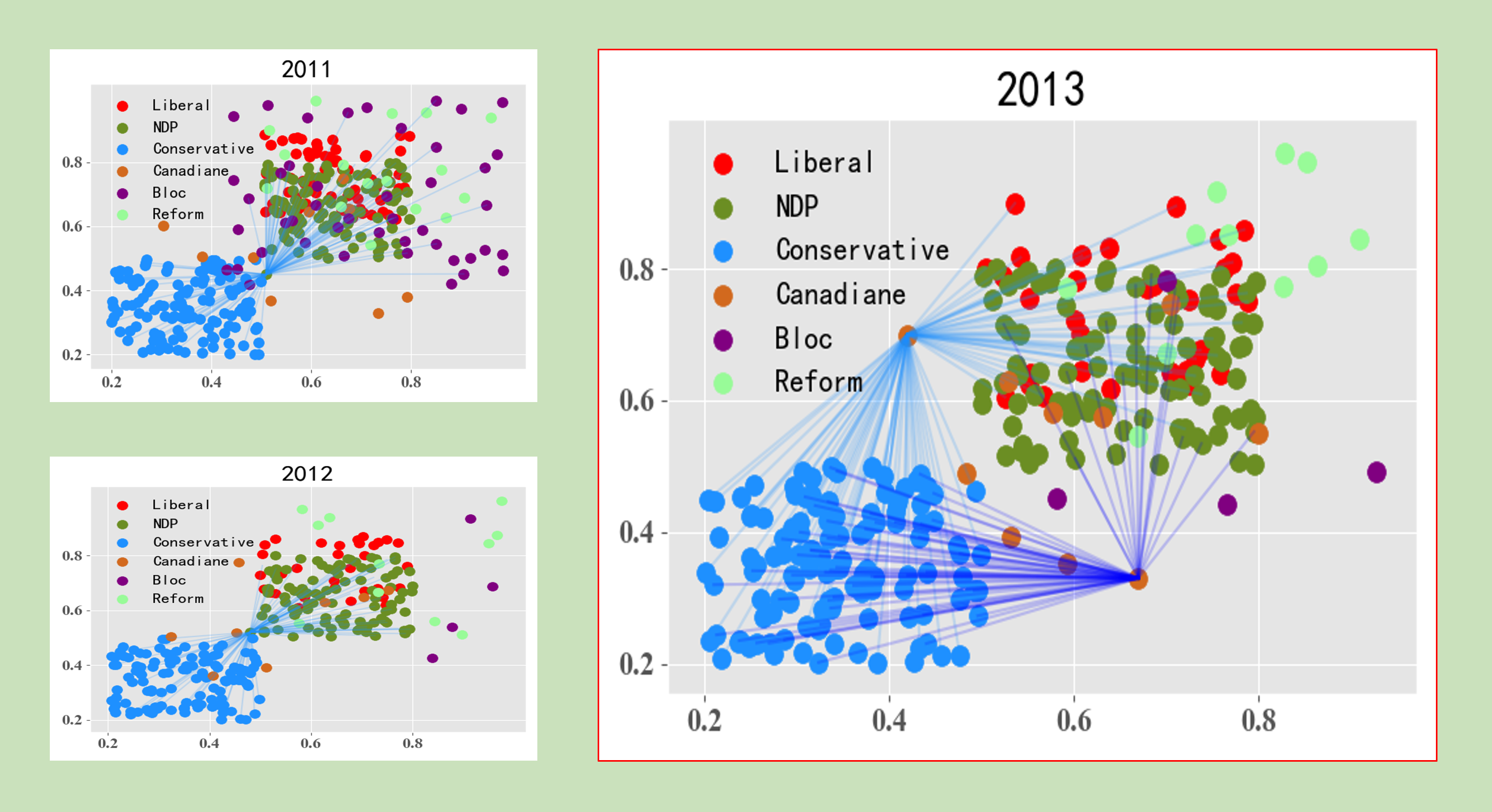}
		\caption{Canadian parliament voting networks. In 2011 and 2012, the most of votes gathered in one node, while the voting network changed in 2013 since there are two aggregation points. It is indicated that 2013 is a change point of the Canadian parliament voting network. }	
		\label{figure1}
	\end{figure}

	Time-varying network has shed a light on the spatio-temporal information of this network (graph), which represents multiple structure of nodes and dynamic changes \cite{wang2020dynamic}. A stable and gradual development of such network is essential to many real-world applications such as city-wide traffic management, weather detection, and detection of epidemic outbreaks \cite{chen2019periodicity,wang2017fast,gong2018network, wang2019deep,chen2021dynamic}. As one of the anomaly detection tasks, change point detection (CPD) concentrates on identifying the anomalous timestamps in temporal networks, which deviates from the normal network evolution \cite{huang2020laplacian}. For example, in the Canadian parliament voting network, there is an anomalous voting pattern in 2013, as shown in the right part of Figure~\ref{figure1} (a). The graph in the red box reveals different voting interactions compared with previous two years (2011 and 2012). CPD can identify anomalies like this from the time-varying network sequences.

\begin{figure}[h]
		\centering
		\includegraphics[width=0.8\linewidth]{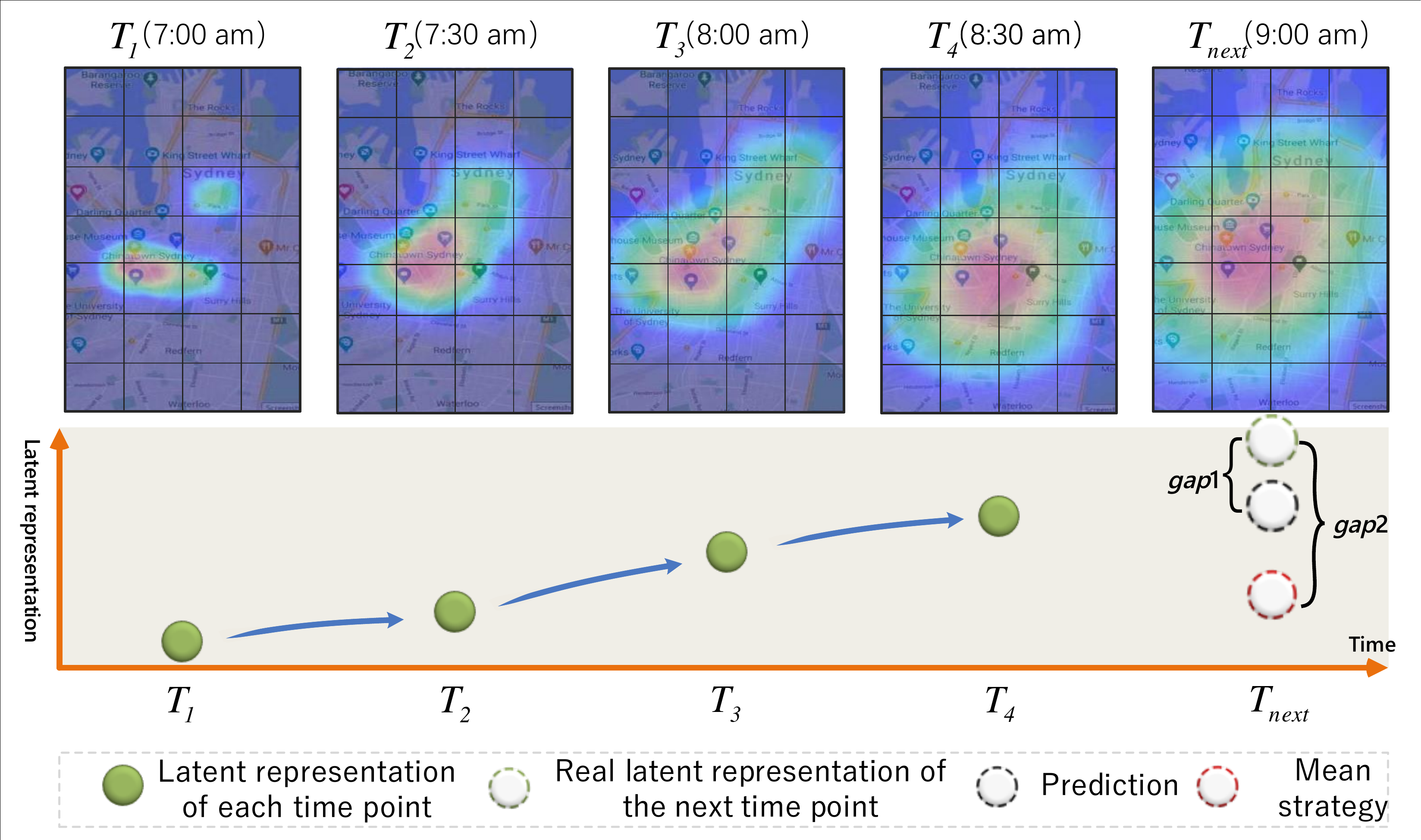}
		\caption{CPD with time-varying prediction strategy.}	
		\label{figure2}
	\end{figure}

	Nowadays, existing methods mainly compare the target network with a learned normal graph pattern \cite{huang2020laplacian, huang2021scalable}. This normal pattern is extracted from previous networks in a given time-window via a mean or global representation learning method~\cite{huang2020laplacian,ide2004eigenspace,koutra2012tensorsplat}. Even these methods have achieved encouraging results on the synthetic test, most of them neglect capturing the real-world dynamic evolution patterns. In other words, they are suitable for implementing in a relatively stable environment \cite{li2014performance}, but not always able to capture anomalies in temporal networks (e.g., traffic network) , as they do not take into account the evolution trend across the temporal networks \cite{chen2020citywide,qu2022forecasting,liu2022traffic}. For example, the upper part of Figure \ref{figure2} illustrates the temporal changes of crowd flow in the Sydney traffic network. If these traffic maps are embedded into a latent space, their latent representations could be presented in the bottom part of this figure. There is a clear upward trend in the real-world traffic network. Considering the change point detection for the next timestamp, we aim to capture the time-evolving trend and make an accurate prediction to the traffic map at $T_{next}$. When a prediction (represented by the black dotted circle) is obtained, we can compare it with the real map representation (the green dotted circle) at $T_{next}$. However, if we use an average or global method to compute the normal graph pattern of previous networks, we can only capture the latent representation as illustrated by the red dotted circle. Apparently, the $gap_2$ is much larger than $gap_1$, suggesting that only extracting the previous normal graph pattern based on the average method leads to a distorted result. Moreover, a time-varying network is not only reflected by the link connections, but also varied in the weight changes. Even though a few of methods can capture the changes for link connections, they do not consider the dynamic weights \cite{wang2021optimal}. 
	
	Motivated by the above issues, we draw inspiration from several latent space prediction model \cite{deng2016latent,gong2020online} and propose a new evolution method for the change point detection. One of the purposes in this paper is to capture the evolution trend of a time-varying network. Once the trend is learned, we can make a prediction of the network with the evolving patterns. Accordingly, the difference between the prediction and the actual network can be recognized as a criteria to identify the change point. After casting the problem as a prediction task, there are two challenges along with that need to be carefully handled.

	\begin{itemize}
		\item One is how to capture temporal dependencies and evolution trends.
		\item The second is how to identify the anomalous network by considering the prediction result.
	\end{itemize}	
	
	To address these challenges, we have developed a prediction model for the time-varying network, and proposed a trade-off strategy based on the prediction result and the previous normal graph pattern to detect change points. Specifically, we first embedded the original network into latent spaces via matrix factorization model. Then, our method goes a step further to capture the dynamic movements of these latent spaces from  current timestamp to the next via the latent transition learning process. When the latent spaces and transition matrices are learned, we can make a prediction to the target network. At second, we choose the Laplacian spectrum strategy to compare the differences between the prediction and the actual network. At last, the normal graph pattern learned from previous networks is taken into consideration as a trade-off strategy to identify whether the target network is a change point. We summarize the main contributions and innovations of this paper as follows:
	
	\begin{itemize}
		\item Unlike existing methods detecting change points directly \cite{wang2017fast,huang2020laplacian,huang2021scalable}, we formulate the CPD problem as a temporal network prediction task, and propose a data-driven forecasting method via the \underline{l}atent \underline{e}volution \underline{m}odel, called LEM-CPD, which learns network evolution trends over time.
		\item To improve the effectiveness and robustness of the model in various real-world scenarios, we further take long-term and periodic properties into consideration, and propose a novel long-term guidance method based on the weighted multi-view learning.
		\item We develop a trade-off strategy to identify change points based on the Laplacian spectrum method, which balances the importance between the
		prediction network and the normal graph pattern extracted from previous networks.
	\end{itemize}
	
	\section{Related Work}
	\subsection{Change Point Detection}
	In this section, we briefly introduce some related literatures of CPD, which can be used to discover outliers in network and sequences \cite{dong2014comparisons,gong2015research,gong2017nspfi,huang2022series}. \cite{peel2015detecting} provided the first attempt on transforming the CPD problem to detect the times of fundamental changes. The proposed LetoChange method used Bayesian hypothesis testing to determine whether a model with parameter changes is acceptable. \cite{wang2017fast} proposed a Markov process based method, which treats the time-varying network as a first order Markov process. This work used joint edge probabilities as the "feature vector".
	\cite{koutra2012tensorsplat} proposed a tensor-based CPD model. Tensor factorization is widely recognized as an effective method to capture a low-rank representation in the time dimension. In this method, after getting the latent attributes of each time point, the clustering and outlier detection methods were used to identify change points. Activity vector method was proposed by \cite{ide2004eigenspace}, it focused on finding anonymous networks which are significant deviation from previous normal ones. Activity vector leveraged the principal eigenvector of the adjacency matrix to present each network, it considered a short term window to capture the normal graph pattern. Currently, \cite{huang2020laplacian} proposed a Laplacian-based detection method for CPD, named LAD. It utilised the Laplacian matrix and SVD to obtain embedding vector of each network. In additional, two window sizes are involved to capture both short and long term properties. \cite{wang2021optimal} proposed an efficient approach for network change point localization. However, this method only considered the changes in link connections that cannot capture the dynamics for weight graphs.
	
	As aforementioned discussion, most of the existing relevant methods did not fully consider the dynamic trends in the time-varying networks when detecting change points.
	%

	\subsection{Dynamic Latent Space Model}
	
	Recent studies have indicated that the latent space learning methods provide excellent solutions for the dynamic network problems, e.g., flow prediction \cite{gong2020online}, community 
	detection systems \cite{zhang2012overlapping}, social network evolution learning \cite{xu2014dynamic}, sequential analysis \cite{dong2018f,dong2018rnsp,qiu2021efficient,gong2021inferring}, and traffic speed prediction \cite{deng2016latent}. The benefits of a latent space model is to obtain the low-rank approximations from the original features, so that obtain a more compact matrix by removing the redundant information of the large scale matrix \cite{li2019sample,li2020field}. \cite{sarkar2007latent} proposed a dynamic model for a static co-occurrence model CODE, where the coordinates can change with time to get new
	observations. There are several advanced investigations focusing on traffic analysis under the latent space assumption. For example, to address the issues of finding abnormal activities in crowded scenes, \cite{thida2013laplacian} devised a spatio-temporal Laplacian eigenmap strategy to
	detect anomalies. \cite{jing2018high} leveraged a tensor structure to resolve the high-order time-varying  prediction problem. The key characteristics of the original time sequences can be represented by the learned latent core tensor. \cite{deng2016latent} developed
	a real-time traffic speed prediction model that learns temporal and topological features from the road network. It then used an online latent space learning method to solve the proposed constrained optimization problem. \cite{wang2011efficient} took the advantage of dynamic latent space learning to address the temporal analysis problem very efficiently on the large-scale streaming data. In other applications, \cite{gong2018network,gong2020online} used the online matrix factorization model to solve the crowd flow distribution problem.
	
	In summary, latent space models can effectively capture dynamic evolving trends from real-world networks, as well as learning the latent embedding for such networks.

	\section{Problem Formulation}
	\label{motivation}

	In the CPD problem, we use a directed graph $\bm G_t = (\bm V_t, \bm E_t)$ to define the network at timestamp $t$, where $\bm E_t$ is the group of edges and $\bm V_t$ denotes the group of vertexes. An edge $e_t(v_{i}, v_{j}, w_{ij})$ denotes the connection between node $v_i$ and node $v_j$ with weight $w_{ij}$ at timestamp $t$. Following the same setting in \cite{huang2020laplacian,koutra2012tensorsplat}, $w$ = 1 for all edges in unweighted graphs and $w$ $\in$ $ \mathbb{R}^+$
	for weighted graphs; and the number of nodes in the graph is assumed to be constant across all timestamps. 
	
	Given a certain time window $T$, our work first aims to predict the next network $ \bm G_{T+1}$ by using a series of previous networks, $\{\bm G_t\}_{t=1}^T$. Second, the next snapshot $\bm G_{T+1}$ will be identified as the change point or not via a trade-off strategy by leveraging the prediction $\bm G_{T+1}$ and the previous normal graph pattern. Among which, the \textbf{normal graph pattern} indicates the average network embedding learned from previous $\{\bm G_t\}_{t=1}^T$ in a given time window. The detailed learning method is presented in Section 4.5.

	\section{Methodology}
	
	\subsection{Temporal Topology Embedding}
	Considering inherent complexities of the time-varying network, we leverage the latent space model to represent dynamic graphs. Latent space models are widely applied into solving network-wide problems, e.g., community detection \cite{zhang2012overlapping}, social and heterogeneous networks construction\cite{wang2020dynamic}, and network-wide traffic prediction\cite{gong2020online,gong2020potential}, etc. The benefit of low-rank approximation of latent space model is to get a more compact matrix by removing the redundant information of the large-scale matrix \cite{zhang2015abnormal}. For a slice $\bm G$ $\in$ $\mathbb{R}^{n \times n}_+$ (the timestamp $t$ is omitted for brevity), we use tri-factorization that decomposes $\bm G$ into three latent representations, in the form $\bm U\bm C \bm V$, where $\bm U$ $\in$ $\mathbb{R}^{n \times k}_+$ and $\bm V$ $\in$ $\mathbb{R}^{k \times n}_+$ denote the latent attributes of start and end nodes respectively, $\bm C$ $\in$ $\mathbb{R}^{k \times k}_+$ denotes the attribute interaction patterns, $n$ is the number of nodes and $k$ is the number of dimension of latent spaces. Therefore, we approximate $\bm G$ by a non-negative matrix tri-factorization form $ \bm U \bm C \bm V$:
	
	\begin{equation}
		\small
		\underset{\bm U,\bm C,\bm V\geq 0}{{ \rm  min}} \; ||\bm G - \bm U\bm C\bm V ||_{F}^{2}.
	\end{equation}
	
	Note that, some tri-factorization methods prefer to use the factorization form $\bm G \approx \bm U\bm C\bm U^{\top}$ because they usually focus on the undirected graph \cite{zhang2012overlapping,pei2015nonnegative}. However, in our problem, some real-world networks are formulated as the directed graph. In this way, $\bm G \approx \bm U\bm C\bm V$ provides a more flexible formulation to represent the graph $\bm G$ \cite{li2009non,wang2011fast}. Thus, our basic problem formulation differs from that of methods in \cite{deng2016latent,gong2018network,gong2020online}.
	
	The reasons that we utilise the non-negativity constraint are: 1) the reasonable assumptions of latent attributes and the better interpretability of the results \cite{li2016graph,luo2019temporal}; 2) all weights in our networks are non-negative, thus latent spaces $\bm U$, $\bm V$ and $\bm C$ should be non-negative as well.
	
	To take the time-series pattern into consideration, we formulate this sequential problem as a time-dependent model. Given a certain time window $T$ (i.e., a set contains $T$ previous networks, and $T$ presents the current time as well), for each timestamp $t$ $\in$ [1, $T$], our method aims to learn the time-dependent latent representations $\bm U_t$, $\bm V_t$ and $\bm C$ from $\bm G_t$. Although the latent attribute matrix $\bm U_t$ is time-dependent, we assume that the latent attribute interaction
	matrix $\bm C$ is an inherent property, and thus $\bm C$ is fixed for all timestamps. This assumption is based on the fact that temporal network exists invariant properties \cite{deng2016latent}. After considering the temporal pattern, our model can be formulated as:
	
	\begin{equation}
		\small
		\underset{\bm U_t,\bm C,\bm V_t\geq 0}{{\rm   min}} \; \mathcal{ J} = \sum_{t=1}^{T} ||\bm G_t - \bm U_t\bm C\bm V_t||_{F}^{2}.
	\end{equation}

	\subsection{Latent Space Evolution}
	\label{section_transition}
	
	Since our model is a time-varying forecasting system, the network is continually changing over time. We not only want to learn the time-dependent latent attributes, but also aim to learn the evolution patterns of latent attributes $\bm U_{t}$ and $\bm V_{t}$ from previous timestamps to the next, i.e., how to learn the evolution forms such as $\bm U_t$ $\rightarrow$ $\bm U_{t+1}$ and $\bm V_t$ $\rightarrow$ $\bm V_{t+1}$. 
	
	To this end, we define two transition matrices $\bm A \in \mathbb{R}_{+}^{k\times k}$ and $\bm B \in \mathbb{R}_{+}^{n\times n}$ to represent the smooth trends of $\bm U_t$ and $\bm V_t$ in \textit{T} previous conditions. The learned transition patterns $\bm A$ and $\bm B$ can be treated as a global evolution trend from previous timestamps. Thus, matrices $\bm A$ and $\bm B$ can be recognized as the \textbf{evolving patterns} of the dynamic network. For example, $\bm A$ approximates the changes of $\bm U$ between time \textit{t}-1 to \textit{t}, i.e., optimizes $\bm U_{t} = \bm U_{t-1}\bm A$. 
	
	Then, let us involve all timestamps in a window-size $T$,  our latent transition learning can be expressed as:
	\begin{equation}
		\small
		\label{L_13}
		\underset{\bm U_{t},\bm V_{t},\bm A,\bm B,\bm C \geq 0}{{\rm min}} \mathcal{T} = \sum_{t=2}^{T}(||\bm U_{t} - \bm U_{t-1}\bm A ||_{F}^{2}+
		||\bm V_{t} - \bm V_{t-1}\bm B ||_{F}^{2}).
	\end{equation} 
	
	When we consider the temporal topology problem jointly, we have:
	\begin{equation}
		\small
		\label{L1}
		\underset{\bm U_{t},\bm V_{t},\bm A,\bm B,\bm C \geq 0}{{\rm min}} \mathcal{L}_1 = \mathcal{ J} + \lambda_{1}\mathcal{T},
	\end{equation} 
	where $\lambda_{1}$ is the regularization parameter.
	
	To this stage, transition matrices $\bm A$ and $\bm B$ can learn the short-term knowledge from previous networks in a window size. Considering that the long-term knowledge is helpful for the CPD and network evolution problems \cite{wang2017fast,gong2020online}, we also take the long-term pattern into consideration to improve the prediction performance in the next section.
	
	\subsection{Long-term Guidance}
	\label{long-term guidance} 
	
	Dynamic networks, such as climate and travel networks, usually have a strong periodic property. Such property is considered as the long-term guidance which has more gradual trends. Because of this point, we can leverage a long-term window to achieve the large-scaled network evolution, and propose a novel long-term guidance to extract periodic and stable information. Given a long-term set of networks $\{\bm G_t\}_{t=1}^{4T}$ (the window size extends from $T$ to 4$T$ for example, then the long-term guidance will consider the states of 4$T$ previous networks), we learn two guide matrices, $\bm U^{\rm lt}$ and $\bm V^{\rm lt}$, to represent the common characteristic of $\{\bm G_t\}_{t=1}^{4T}$: 
	
	\begin{equation}
		\small
		\label{lt}
		\underset{\bm U^{\rm lt},\bm V^{\rm lt} \geq 0}{{\rm min}} \sum_{t=1}^{4T}(||r_t \odot (\bm G_{t} - \bm U^{\rm lt}\bm V^{\rm lt}) ||_{F}^{2},
	\end{equation} 
	where $r_t$ is the adaptive weight calculated by function: $r_t = \frac{e^{s_t}}{\sum^{4T}_{t=1} e^{s_t}}$, and $s_t$ is the similarity between $\bm G_t$ and the current network $\bm G_{4T}$. We use the matrix L$_2$-norm to calculate the similarities among networks, and then normalize them satisfying $\sum^{4T} s_t =1$. The assumption behind this is that $\bm G_t$ can provide more useful information if it is more similar to $\bm G_{4T}$. $\odot$ is the Hadamard product operator. 
	
	The main idea of the above process is analogous to the multi-view learning which projects multiple views into shared common spaces \cite{zhao2017multi,huang2022series}. The adaptive weight $r_t$ controls how much information can be extracted from $\bm G_t$. Such technique has been successfully used in common space learning \cite{gongspatial, gong2021missing}. The learning approaches of Eq. (\ref{lt}) are:
	
	\begin{equation}
		\bm	U^{\rm lt} = \bm U^{\rm lt} \odot \frac{\sum_{t=1}^{4T}(r_t\odot \bm G_t )\bm V^{\top} }{\sum_{t=1}^{4T}[r_t\odot(\bm U^{\rm lt}\bm V^{\rm lt})]\bm V^{\rm lt}},
	\end{equation}
	
	\begin{equation}
		\bm	V^{\rm lt} = \bm V^{\rm lt} \odot \frac{\sum_{t=1}^{4T}\bm U^{\rm lt \top}(r_t\odot \bm G_t ) }{\sum_{t=1}^{4T}\bm U^{\rm lt \top}[r_t\odot(\bm U^{lt}V^{\rm lt})]}.
	\end{equation}

	After having $\bm U^{\rm lt}$ and $\bm V^{\rm lt}$, the strategy of the long-term guidance is formulated as:
	
	\begin{equation}
		\small
		\label{L_h}
		\underset{\bm U_{T},\bm V_{T},\bm A,\bm B \geq 0}{{\rm min}} \mathcal{H} = ||\bm U^{\rm lt} - \bm U_{T}\bm A||_{F}^{2}+||\bm V^{\rm lt} - \bm V_{T}\bm B ||_{F}^{2},
	\end{equation} 
	where $\bm U_{T}$ and $\bm V_{T}$ denote the latent representations at current timestamp $T$. 
	
	Taking all techniques jointly into consideration, the final optimization function is expressed as: 
	
	\begin{equation}
		\small
		\label{L_final}
		\underset{\bm U_{t},\bm V_{t},\bm A,\bm B,\bm C \geq 0}{{\rm min}} \mathcal{L} = \mathcal{J} + \lambda_{1}\mathcal{T} +  \lambda_{2}\mathcal{H},
	\end{equation} 
	where $\lambda_{2}$ is the regularization factor.

	\textbf{\textit{Make a Prediction.}} The learned latent space matrices $\bm U_{T}$ and $\bm V_{T}$, and the evolving patterns $\bm A$, $\bm B$ and $\bm C$ can be used to make the prediction after solving Equation (\ref{L_final}). Then we have the predicted network $\hat{\bm G}_{T+1}$ as:
	
	\begin{equation}
		\small
		\label{equ_pre}
		\hat{\bm G}_{T+1} = (\bm U_{T}\bm A)\bm C(\bm V_{T}\bm B).
	\end{equation}

	\subsection{Learning Process}
	\label{learn OLS-AO}

	Due to the non-convexity of Eq. (\ref{L_final}), we use an effective gradient descent approach with the multiplicative update method \cite{lee2001algorithms} to find its local optimization. 
	
	\paragraph{Theorem 1} \textit{$\mathcal{L}$ is non-increasing by optimizing $\bm U_t$, $\bm V_{t}$, $\bm A$ $\bm B$ and $\bm C$ alternatively. }

	\begin{small}
		\begin{equation}
			\label{update_U}
			\bm	U_{t} =  \bm U_{t}\odot \frac{\bm G_{t}(\bm C\bm V_t)^{\top}+\lambda _{1}(\bm U_{t-1}\bm A+\bm U_{t+1}\bm A^{\top})+\hat{\lambda _{2}}\bm U^{\rm lt}\bm A^{\top}}{(\bm U_{t}\bm C\bm V_{t})(\bm C\bm V_t)^{\top}+\lambda _{1}(\bm U_{t}+\bm U_{t}\bm A\bm A^{\top})+\hat{\lambda_{2}}\bm U_{T}\bm A\bm A^{\top}},
		\end{equation}
	\end{small}

	\begin{small}
		\begin{equation}
			\label{update_V}
			\bm V_{t} =  \bm V_{t}\odot \frac{(\bm U_{t}\bm C)^{\top}\bm G_{t}+\lambda _{1}(\bm V_{t-1}\bm B+\bm V_{t+1}\bm B^{\top})+\hat{\lambda _{2}}\bm V^{\rm lt}\bm B^{\top}}{(\bm U_{t}\bm C)^{\top}(\bm U_t\bm C\bm V_t)+\lambda _{1}(\bm V_{t}+\bm V_{t}\bm B\bm B^{\top})+\hat{\lambda_{2}}\bm V_{T}\bm B\bm B^{\top}},
		\end{equation}
	\end{small}
	\textit{ $\hat{\lambda _{2}}$ is given by:}

	\begin{equation}
		\label{lambda_2}
		\small
		\hat{\lambda _{2}} = \left\{\begin{matrix}
			\lambda _{2}, \;  t=T \\\;\;\;\;\;0, \;\rm otherwise
		\end{matrix}\right.
	\end{equation}
	
	\begin{equation}
		\small
		\label{update_A}
		\bm A = \bm A\odot \frac{\lambda _{1}\sum_{t=1}^{T}\bm U_{t-1}^{\top}\bm U_{t}+\lambda _{2}\bm U_{T}^{\top}\bm U^{lt}}{\lambda _{1}\sum_{t=1}^{T}\bm U_{t-1}^{\top}(\bm U_{t-1}\bm A)+\lambda_{2}\bm U_{T}^{\top}(\bm U_{T}\bm A)},
	\end{equation} 
	
	\begin{equation}
		\small
		\label{update_B}
		\bm B = \bm B\odot \frac{\lambda _{1}\sum_{t=1}^{T}\bm V_{t-1}^{\top}\bm V_{t}+\lambda _{2}\bm V_{T}^{\top}\bm V^{lt}}{\lambda _{1}\sum_{t=1}^{T}\bm V_{t-1}^{\top}(\bm V_{t-1}\bm B)+\lambda_{2}\bm V_{T}^{\top}(\bm V_{T}\bm B)},
	\end{equation} 
	
	\begin{equation}
		\small
		\label{update_C}
		\bm C = \bm C\odot \frac{\sum_{t=1}^{T}\bm U_t^{\top}\bm G_t\bm V_t^{\top}}{\sum_{t=1}^{T}\bm U_t^{\top}(\bm U_t\bm CV_t)\bm V_t^{\top}},
	\end{equation} 
	where $X^{\top}$ is the transpose of $X$. The proof of \textbf{Theorem 1} and the detailed derivations are deferred to the Appendix. 
	
	Note that, update rules Eq. (\ref{update_U}) - (\ref{update_V}) indicate a time-related learning process, the information from previous and next timestamps ($\bm U_{t-1}$ and $\bm U_{t+1}$) affect each other. This chain updating rules make sure all variables are learned mutually.  
	
	%
	%

	\subsection{Change Point Detection}
	
	By computing the prediction network $\hat{\bm G}_{T+1}$ for timestamp $T$+1, we can compare it with the actual network $\bm G_{T+1}$ to check whether $\bm G_{T+1}$ is a change point or not. Specifically, we use the Laplacian-spectrum-based method which has shown a state-of-the-art performance in graph comparisons \cite{ wilson2008study}. The eigenvalues of the Laplacian matrix capture the structural properties of the corresponding graph. Note that real-world applications contain both undirected and directed graphs. For the undirected graphs, we construct a normalized Laplacian matrix $\bm L_t$, defined as $\bm L_t = \bm I- \bm D_t^{-1/2}\bm W_t\bm D_t^{-1/2}$, where $\bm I$ is the identity matrix; $\bm W_t$ is the weight matrix of $\bm G_t$ and $\bm D_t$ is a diagonal matrix $\bm D_{t,(i;i)}$ = $\sum_j(\bm G_{t,(i;i)})$. 
	For the directed graphs, we leverage the method proposed in \cite{chung2005laplacians} to build the Laplacian matrix:
	\begin{equation}
		\small
		\bm P_{i,j} = \frac{w_{i,j}}{\sum _j w_{i,j}},
	\end{equation}
	
	\begin{equation}
		\small
		\bm L = \bm I - (\bm \phi^{1/2}\bm P \bm \phi^{-1/2} + \bm \phi^{-1/2} \bm P^T \bm \phi^{1/2} ) / 2,
	\end{equation} 
	where $\bm \phi$ is the matrix with the Perron vector \cite{pillai2005perron} of $\bm P$ in the diagonal; $\bm P$ is a transition probability matrix \cite{chung2005laplacians}.
	
	Given the Laplacian matrices for all timestamps, i.e., $\bm L_1$, $\cdots$, $\bm L_{T+1}$ and $\hat{\bm L}_{T+1}$, where $\bm L_{T+1}$ is built by actual network $\bm G_{T+1}$; $\hat{\bm L}_{T+1}$ is built by the prediction $\hat{\bm G}_{T+1}$. We follow a well-performed metric that utilises the singular values calculated of Laplacian matrices $\bm L_t$ as the network attributes \cite{huang2020laplacian}. Then, the vector made up of singular values is denoted as $\overrightarrow{\bm \sigma}_t$, $t$ $\in$ [1, $T$+1]. Let $\overrightarrow{\bm \sigma}_{a}$  ($\overrightarrow{\bm \sigma}_{T+1}$) denote the vector of actual network,  $\overrightarrow{\bm \sigma}_{p}$ (singular values of $\hat{\bm L}_{T+1}$) denote the vector of prediction network. The first anomaly score $\bm Z_1$ based on the cosine distance is defined as:
	\begin{equation}
		\small
		\label{Z-1}
		\bm Z_1 = 1-\frac{\overrightarrow{\bm \sigma} ^{\top}_{p}\overrightarrow{\bm \sigma}_{a}}{||\overrightarrow{\bm \sigma}_{p}||_2||\overrightarrow{\bm \sigma}_{a}||_2}.
	\end{equation}
	
	It is evident that the more dissimilar between the prediction and the actual network, $\bm Z_1$ is closer to 1, and vice versa\footnote{The range of $\bm Z_1$ is [0,1] because of the non-negativity of $\overrightarrow{\bm \sigma}$}.
	
	\paragraph{Trade-off Strategy} The score of $\bm Z_1$ reflects the difference between the prediction and ground truth. This evaluation is more suitable for the temporal network that evolves gradually. However, the normal graph pattern extracted from the previous networks is also important since it can reflect the average evolution representation. Thus we introduce the second anomaly score $\bm Z_2$ which considers the normal graph patterns. Let $\overrightarrow{\bm \sigma}_{nor}$ be the average vector of previous snapshots, i.e., $\overrightarrow{\bm \sigma}_{nor}$ = ($\sum_{t=1}^T \overrightarrow{\bm \sigma}_{t})/T $, $\bm Z_2$ is defined as:
	\begin{equation}
		\small
		\label{Z-2}
		\bm Z_2 = 1-\frac{\overrightarrow{\bm \sigma} ^{\top}_{nor}\overrightarrow{\bm \sigma}_{a}}{||\overrightarrow{\bm \sigma}_{nor}||_2||\overrightarrow{\bm \sigma}_{a}||_2}.
	\end{equation}
	
	Finally, the combined anomaly score $\bm Z$ is a trade-off between $\bm Z_1$ and $\bm Z_2$:
	\begin{equation}
		\label{Z-final}
		\small
		\bm Z = \alpha \bm Z_1 + (1- \alpha \bm Z_2),
	\end{equation}
	where $\alpha$ is the trade-off factor. Algorithm 1 summarizes our learning and detection process in LEM-CPD.

	\subsection{Complexity Analysis}
	\label{time complex}
	Here we analyse the time complexity of LEM-CPD algorithm. After the prediction process, the time complexity of the change point detection is mainly affected by SVD. Even though randomized SVD has an $O(n^2k)$ complexity, it is not involved in the update loop of variables (such as $\bm U_t$ and $\bm V_t$). In the prediction process, 
	the time complexity of Eq. (\ref{update_U}) - Eq. (\ref{update_V}) is dominated by the matrix multiplication operations in each iteration. Therefore, for each iteration, the computational complexity of LEM-CPD is $O(Tkn^2)$. Compared with the latest CPD method LAD  ($O(n^2k)$) \cite{huang2020laplacian}, the time complexity of our method is $O(Tkn^2)$, which is higher than LAD.

	\begin{algorithm}[]
		\small
		\caption{LEM-CPD}
		\LinesNumbered 
		\KwIn{network matrices [$G_{1},\cdots,G_{T}$]; actual network $G_{T+1}$; long-term guidanc $U^{lt}$, $V^{lt}$.}
		\KwOut{$Z$ score.}
		initialize $U_{t}$, $V_{t}$, $A$, $B$ and $C$ by previous detection values.\\ 
		\For{\textit{Iter} = 1 to \textit{Max}}{
			\eIf{ $\left | \mathcal{L}_i - \mathcal{L}_{i+1} \right |$ / $\mathcal{L}_i$ $\geq$ $\varepsilon $}{
				\For{\textit{t} = 1 to \textit{T}}{
					update $U_{t}$ and $V_{t}$ \textbf{By} Equations (\ref{update_U}) and (\ref{update_V}).\\
					
				}	
				update $A$, $B$ and $C$ \textbf{By} Equations (\ref{update_A}) - (\ref{update_C}).\\		
			}{
				Break
			}
		}
		
		Prediction network $\hat{G}_{T+1}$ = $(U_{T}A)C(V_{T}B)$.\\
		Calculate  $L_1$, $\cdots$, $L_{T+1}$ and $\hat{L}_{T+1}$.\\
		get $\overrightarrow{\sigma}_{t}$; $\overrightarrow{\sigma}_{a}$; $\overrightarrow{\sigma}_{p}$ and $\overrightarrow{\sigma}_{nor}$.\\
		Calculate $Z$ score \textbf{By} Equations (\ref{Z-1}) - (\ref{Z-final}).
	\end{algorithm}

	\section{Experiments and Results}
	\label{experiment}
	\paragraph{Baselines}
	We compare the proposed method LEM-CPD with the following baselines. All parameters of the proposed method and baselines are optimized by the grid search method in the prepared validation dataset. For example, in the real-world Flow dataset, five workdays in one week is chosen as the validation data, then we test our method on the next week. 

	\begin{itemize}
		\item \textbf{LT-A}. We compared the next network with its long-term average network.
		
		\item \textbf{LAD}. \cite{huang2020laplacian} proposed a Laplacian-based detection method for CPD. It used the Laplacian matrix and SVD to obtain embedding vector of each network. In additional, two window sizes are involved to capture both short and long term properties.
		
		\item \textbf{LADdos}. \cite{huang2021scalable} devised a variant method of LAD to identify time steps where the graph structure deviates significantly
		from the norm. 
		
		\item \textbf{EdgeMonitoring}. This method used joint edge probabilities as the "feature vector", and formulated the time-varying network as a first order Markov process \cite{wang2017fast}.
		
		\item \textbf{TensorSplat}. \cite{koutra2012tensorsplat} proposed a tensor-based CPD model, which utilised the CP decomposition to capture a low-rank space in the time dimension. Then an outlier detection method is used to identify the change points.
		
		\item \textbf{Activity vector}. This algorithm used the principal eigenvector of the adjacency matrix to present each network and only considered short term properties \cite{ide2004eigenspace}.	
		
		\item \textbf{LEM-CPD$_{-lt}$}. The variant of the proposed method, which removes the long-term guidance.
	\end{itemize}
	
	\paragraph{Metrics}
	
	The Hit Ratio (HR@\textit{K}) metric is a popular evaluation method that reflects the proportion of right detected anomalies in the top \textit{K} most anomalous points. Unlike the synthetic experiments, we cannot get the ground truth in the real-world datasets, and thus the well-known anomalous timestamps are used as labels.
	
	\subsection{Synthetic Experiments}
	\label{experiment_Synthetic}
	\textbf{Setting:} We chose a widely used data generator Stochastic Block Model (SBM) \cite{holland1983stochastic} to 
	generate synthetic data. We followed two settings in the data generation process in \cite{huang2020laplacian}. 1) \textbf{Pure datasets}: four time-varying networks with 500 nodes and 151 time points are generated, which contain 3, 7, 10, 15 change points respectively. Therefore, we can use HR@K, where K is setting to 3, 7, 10, 15 to evaluate all competitive methods.  \textbf{Hybrid dataset}: we generate four homogeneous networks of the Pure datasets, while contains both change points and event points according to the hybrid setting in \cite{huang2020laplacian}. Thus, HR@K (K = 3, 7, 10, 15) is also suitable in this dataset. In the experiments, our short and long term window sizes are set to 3 and 12 respectively; the change point detection threshold is set to 0.5; the trade-off factor $\alpha$ is set to 0.2. 
	
	\textbf{Discussion:} Table \ref{table_synthetic} presents the HR@K accuracy of all test methods on eight datasets. It is apparent that there are several methods which can effectively handle the CPD problem in the synthetic datasets, e.g., LEM-CPD (ours), LAD, and EdgeMonitoring. It is mainly because the normal evaluation pattern in the synthetic datasets is very stable, so that these method can easily detect anomalies if some dramatically changes appeared. TensorSplat method cannot capture all change points in the sparse datasets. Unlike LADdoc, LAD and LEM-CPD, Activity vector chooses the principal eigenvectors of the adjacency matrix to represent the graph embedding directly, the latent space may not be learned well compared with the Laplacian spectrum method. Furthermore, as the results shown in Table \ref{table_synthetic}, we can see that without the long-term guidance, LEM-CPD$_{-lt}$ performs
	worse than our final model, which demonstrates the effectiveness of the long-term guidance. For methods that use the prediction process, such as LEM-CPD$_{-lt}$ and LEM-CPD, can achieve the best detection accuracy as shown in Table \ref{table_synthetic}.
	
\begin{figure}[t]
		\centering
		\includegraphics[width=1\linewidth]{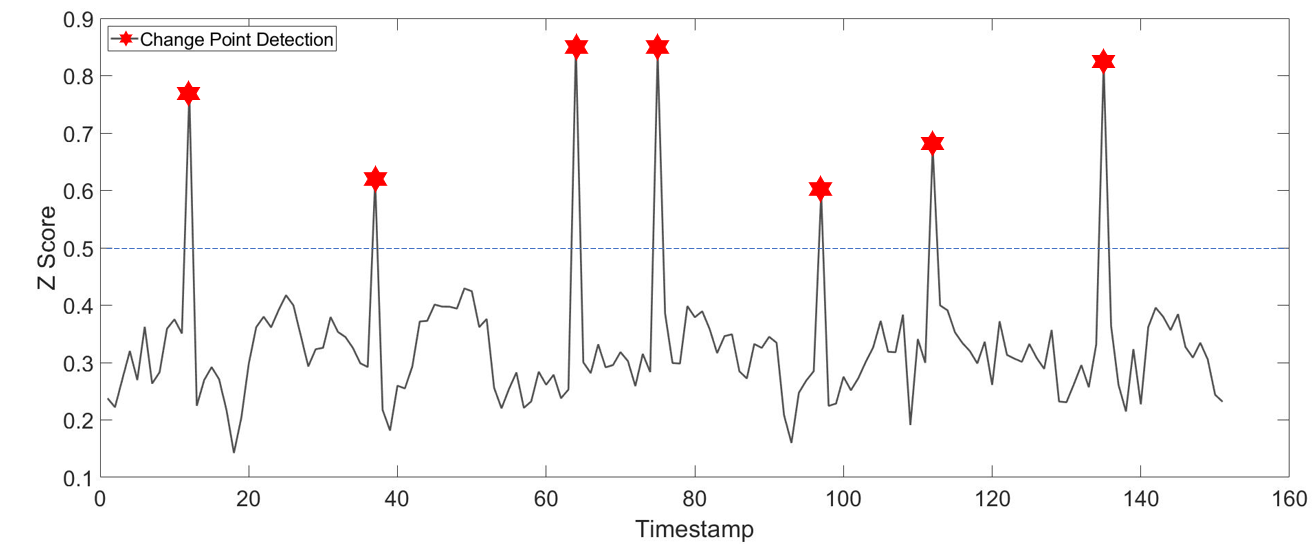}
		\caption{LEM-CPD correctly identifies the change points in the Pure dataset.}	
		\label{result-1}
	\end{figure}

	Figure \ref{result-1} provides the detection results and their $\bm Z$ scores in the Pure dataset with seven change points. The seven most anomalous points are correctly detected. We can clearly see that the $\bm Z$ scores of change points are markedly different from the normal ones. The scenario proves our point that the normal evolutionary pattern of the synthetic network is stable. Therefore, in our trade-off strategy, the score $\bm Z_2$ can greatly handle this situation, so that we use a small $\alpha$ in the experiments.

	\begin{table*}[]
		\centering
		\caption{The Performance on the synthetic datasets.}
		\label{table_synthetic}
		\renewcommand\arraystretch{1}
		\setlength{\tabcolsep}{1mm}
		\begin{tabular}{l|ccccccccc}
			\hline
			Metric          & \multicolumn{2}{c}{HR@3} & \multicolumn{2}{c}{HR@7}& \multicolumn{2}{c}{HR@10} & \multicolumn{2}{c}{HR@15}& Average \\
			Dataset         & Pure        & Hybrid  & Pure        & Hybrid& Pure        & Hybrid& Pure        & Hybrid     \\ \hline
			LT-A    & 66.7\%      & 33.3\%  & 42.9\%      & 28.6\%& 40.0\%      & 20.0\%& 46.7\%      & 33.3\%  &  38.9\%  \\
			TensorSplat   \cite{koutra2012tensorsplat}  & 66.7\%      & 66.7\%  & 71.4\%      & 42.9\%& 70.0\%      & 60.0\%& 73.3\%      & 66.7\%  &  64.7\%    \\
			Activity vector \cite{ide2004eigenspace}& 66.7\%      & 33.3\%  & 57.1\%      & 14.2\%& 60.0\%      & 30.0\%& 66.7\%      & 33.3\%       &  45.2\%\\
			EdgeMonitoring \cite{wang2017fast} & 100.0\%       & 66.7\%   & 71.4\%       & 57.1\% & 50.0\%       & 60.0\% & 73.3\%       & 80.0\%  &  69.8\%    \\
			LAD        \cite{huang2020laplacian}     & 100.0\%       & 66.7\%   & 100.0\%       & 85.7\% & 90.0\%       & 80.0\% & 86.7\%       & 86.7\%  &  86.9\%  \\ 
			LADdoc        \cite{huang2021scalable}     & 100.0\%       & 66.7\%   & 100.0\%       & 85.7\% & 80.0\%       & 90.0\% & 80.0\%       & 86.7\% & 86.1\%   \\ \hline
			LEM-CPD$_{-lt}$  (Ours)       & 100.0\%      & 100.0\%   & 100.0\%       & 85.7\% & 90.0\%       & 80.0\% & 80.0\%       & 86.7\%  &90.3\%     \\ 
			LEM-CPD  (Ours)       & 100.0\%       & 100.0\%  & 100.0\%       & 85.7\% & 90.0\%       & 90.0\% & 86.7\%       & 86.7\%  & 92.4\%    \\ \hline
		\end{tabular}
	\end{table*}

	\begin{table*}[]
		\centering	
		\caption{The performance on the real-world datasets.}
		\label{table_real-world}
		\setlength{\tabcolsep}{2mm}
		\renewcommand\arraystretch{1}
		\begin{tabular}{l|ccccccc}
			\hline
			Metric           & \multicolumn{2}{c}{Flow}  & \multicolumn{2}{c}{Maintenance}  & \multicolumn{2}{c}{Senate}  & Average\\
			Dataset          & HR@3 & HR@6 & HR@1 & HR@3 & HR@1 & HR@2 \\ \hline
			LT-A        & 23.3\%  & 16.7\%  &  100.0\% &  66.6\%  &  50.0\%  & 50.0\% & 51.1\%\\
			TensorSplat \cite{koutra2012tensorsplat}       & 50.0\%  & 43.3\%   & 100.0\% & 66.6\% & 100.0\% &50.0\% & 68.3\% \\
			Activity vector \cite{ide2004eigenspace}     & 33.3\%  & 26.7\%    & 0.0\% & 33.3\% & 50.0\%&  50.0\% &32.2\% \\
			EdgeMonitoring  \cite{wang2017fast}    & 53.3\%    &  40.0\%     & 100.0\% & 66.6\%  & 100.0\%& 100.0\% & 76.6\%\\
			LAD      \cite{huang2020laplacian}           & 40.0\%  & 36.7\%   & 100.0\% & 33.3\% &100.0\% & 100.0\% & 68.3\%\\
			LADdoc       \cite{huang2021scalable}            & 53.3\%  & 40.0\%   & 100.0\% & 33.3\% &100.0\% & 100.0\%
			& 71.1\% \\
			\hline
			LEM-CPD$_{-lt}$ (Ours)       & 76.6\%  &63.3\%    & 100.0\% & 66.6\% & 100.0\% & 100.0\% & 84.4\% \\
			LEM-CPD   (Ours)           & 90.0\%  & 76.6\%  &100.0\% & 66.6\%  & 100.0\%& 100.0\% & 88.9\%   \\ \hline
		\end{tabular}
		
	\end{table*}
	
	\subsection{Real-world Experiments}
	\label{experiment_real}
	\textbf{Setting:} To evaluate the performance of our method in the real-world time-varying networks, we conduct experiments on three datasets. The selection of K of HR@K is based on the number of change points in the corresponding dataset. For example, there are two change points in the Senate dataset, then K will be chosen from 1 and 2. 
	
	1) \textbf{Flow dataset}: a large-scale, real-world dataset that is collected from
	the city transport. This data records the crowd flows among the entire city train network. In a weekday, there exist six change points during the rush and non-rush transition period. 2) \textbf{Maintenance dataset}: The Maintenance of metro lines will change the connections among stations. In this dataset, the one line of the city train network had been maintained on three separated days. Therefore, three change points are  included in the dataset (21 days in total). 3) \textbf{Senate dataset}: a social connection network between legislators during the 97rd-108th Congress \cite{fowler2006legislative}. In this dataset, 100th and 104th congress networks are recognized as the change points in many references \cite{wang2017fast,huang2020laplacian}. The performances of different methods
	are summarized in Table \ref{table_real-world}.

	\textbf{Discussion:} Our model, LEM-CPD significantly outperforms all other comparative methods as shown in Table 2, especially LEM-CPD has outperformed in the Flow dataset. In this test, we report the average accuracy of HR@3 and HR@6 in five weekdays. Compared with LAD and Activity vector, the prediction strategy can capture much more dynamic trends with time evolving. As the motivation illustrated in the Introduction, LADdoc, LAD and Activity vector may difficult to identify the natural gaps between the previous normal graph pattern and the actual network. TensorSplat and EdgeMonitoring achieved the second and third positions because they considered a part of evolution trend during their learning process, e.g., tensor-based methods are able to get the evolution information. In the real-world experiments, we set up a larger $\alpha$=0.6 to let the prediction score participate more in the final decision. There presents similar conclusions in the Maintenance dataset because the city train network has the strong  dynamic characters. However, in the Senate dataset, LADdoc, LAD and EdgeMonitoring methods can also get 100\% accuracy, it is because the 100th and 104th congress networks are significant deviation from others.

	\begin{table}[]
		\centering
		\caption{The prediction comparisons on the Flow dataset.}
		\setlength{\tabcolsep}{4mm}
		\begin{tabular}{c|c|c}
			
			\hline
			& Weekdays & Weekends \\ \hline
			GPR        & 3.283    & 2.333    \\
			LSM-RN-All & 3.713    & 2.105    \\
			SARIMA     & 1.939    & 2.270    \\
			HA         & 1.652    & 2.176    \\
			OLS \cite{gong2020online}       & \textbf{1.531}    & \textbf{1.944 }   \\ \hline
			LEM-CPD (Ours)       & \underline{1.615}    & \underline{2.037 }   \\ \hline
		\end{tabular}
		\label{table_predition}
	\end{table}
	

	\subsection{Prediction Evaluation and Visualization}

	\subsubsection{Baselines and Metric used in the Prediction Evaluation}
	
	$\bullet$ \textbf{HA:} We predict CFD by the historical average method on each
	timespan. For example, all historical time spans from
	9:45 AM to 10:00 AM on Tuesdays are utilized to do the forecast for
	the same time interval.
	
	$\bullet$ \textbf{OLS:} the latest Online latent space based model utilizing side information \cite{gong2018network}.

	$\bullet$ \textbf{LSM-RN-All:}  A state-of-the-art matrix factorization based method to predict network-wide traffic speed problem \cite{deng2016latent}.
	
	$\bullet$ \textbf{SARIMA:} A linear regression model with seasonal property to effectively predict future values in a time series.
	
	$\bullet$ \textbf{GPR:} Gaussian process regression (GPR) would handle the spatiotemporal pattern prediction in a stochastic process. It usually suffers from a heavy computational cost \cite{rasmussen2003gaussian}.

	\textbf{Measures.} The metric used in this paper is Mean Absolute Error (MAE), as it is generally employed in evaluating time series accuracy \cite{gong2020network,qu2022forecasting}.
	
	\begin{displaymath}
		MAE = \frac{\sum_{i=1}^{\Omega}|c_{i}-\hat{c}_{i}|}{\Omega},
	\end{displaymath}
	where $\hat{c}_{i}$ is a forecasting value and $c_{i}$ is the ground truth; $\Omega$ is the number of predictions.

	\begin{figure*}[t]
		\centering
		\subfloat[Timestamp: 8:00 AM - 8:15 AM.]{\includegraphics[width=1\linewidth]{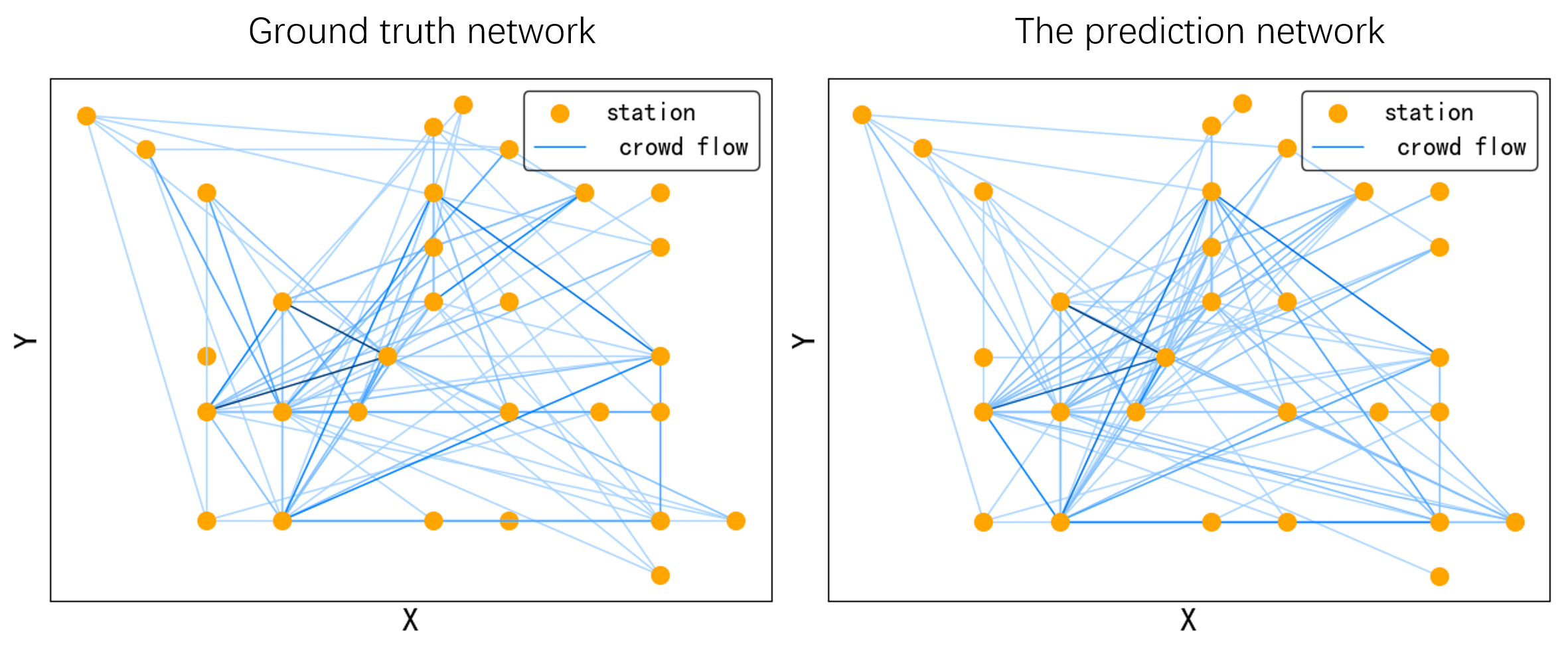}} \\
		\subfloat[Timestamp: 17:00 PM - 17:15 PM.]{\includegraphics[width=1\linewidth]{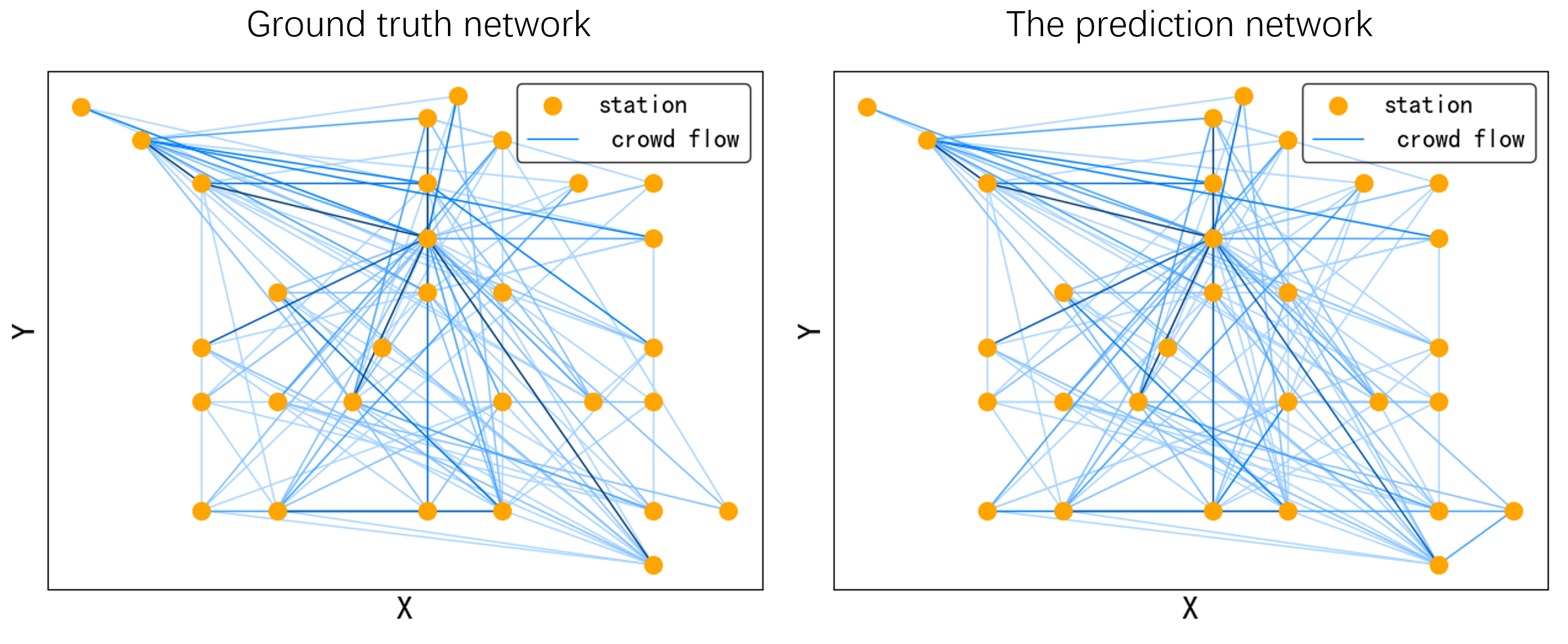}} 
		\caption{The visualization of our prediction.} 
		\label{fig-pred}
	\end{figure*}

	\subsubsection{Results and Visualization}

	This part of experiments is designed to assess the prediction ability of our model. Table \ref{table_predition} reports the average prediction error (mean absolute error) on the Flow dataset. We compared our model with some state-of-the-art methods. In the prediction process, LEM-CPF can achieve the second best result because the best model OLS involves much more external information. 
	
	Figure \ref{fig-pred} gives an intuitive presentation of the flow prediction in Flow dataset; we pick up 30 stations to keep the image clear. Figure \ref{fig-pred} (a) and (b) illustrate the ground truth and prediction of the crowd flow network, each node presents a station, the weight $w_{ij}$ $>$ 0 if there exists a passenger flow in the current timestamp. The depth of color indicates the number of crowd flows. It is apparent that our prediction can construct the flow network precisely.

	\subsection{Sensitivity of Parameters}

	
	\begin{figure}
		
		\centering
		\subfloat[Trade-off Factor $\alpha$.]{\includegraphics[width=0.35\linewidth]{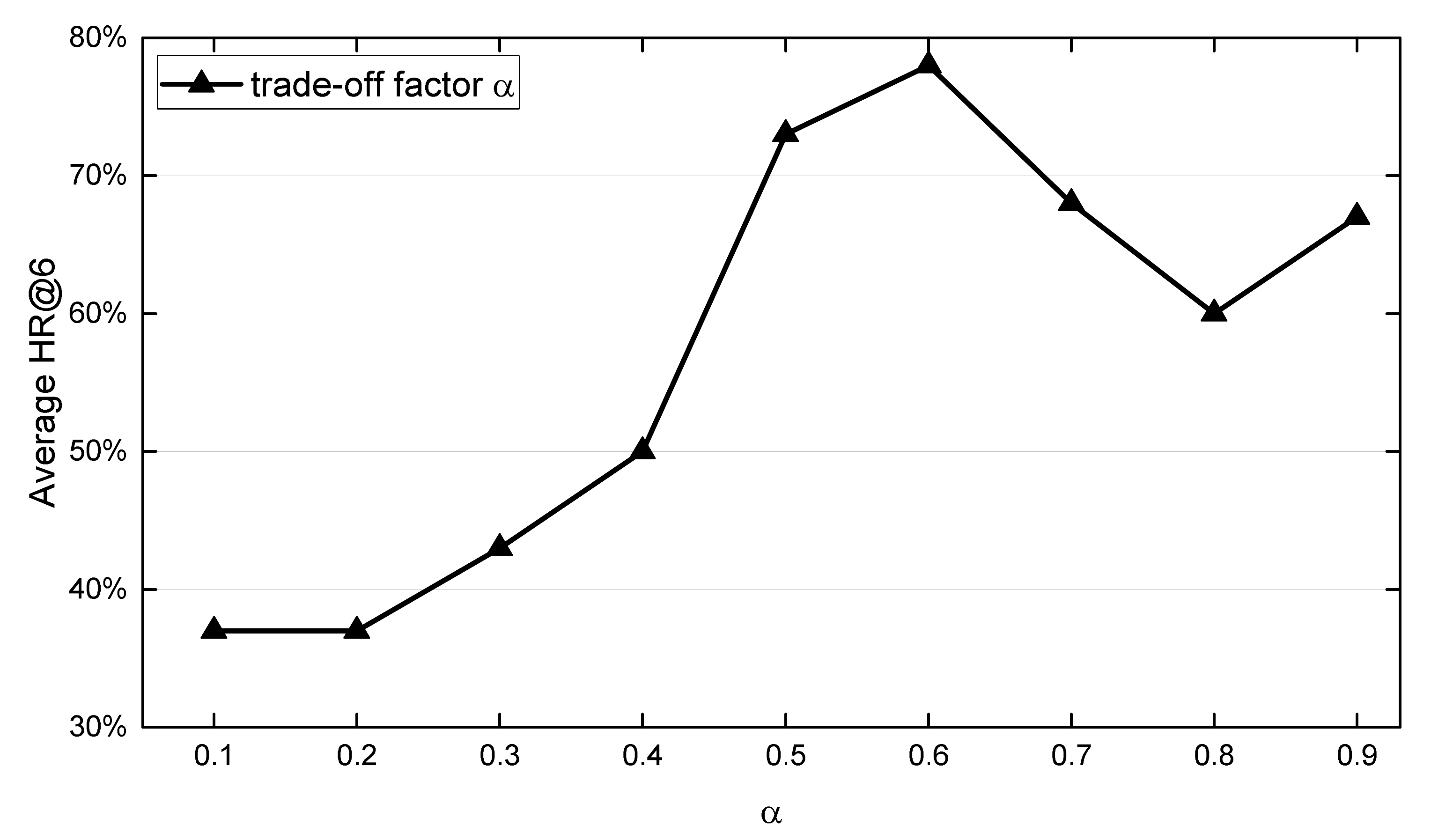}} 
		\subfloat[Factor $\lambda_{1}$.]{\includegraphics[width=0.35\linewidth]{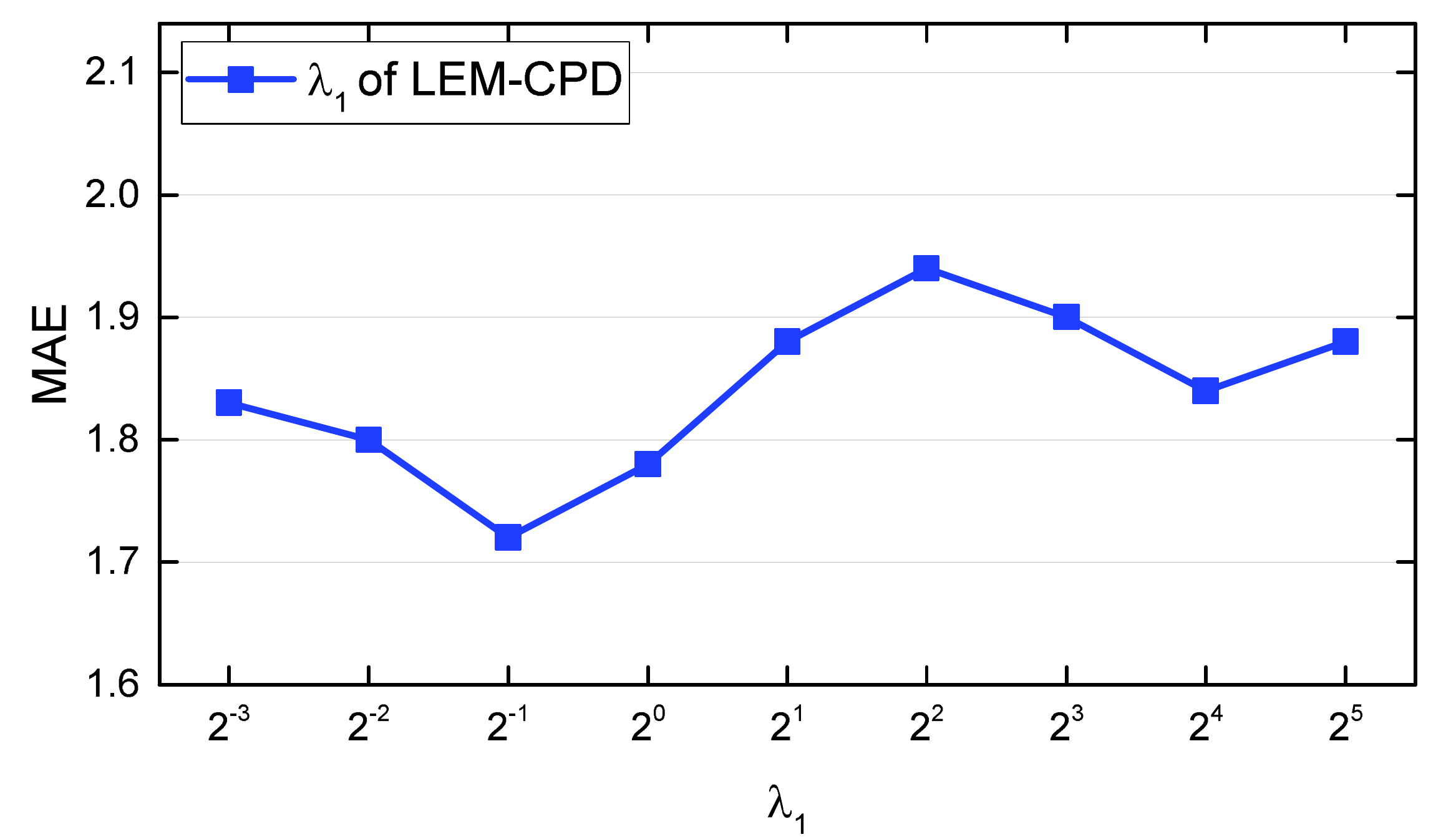}} 
		\subfloat[Factor $\lambda_{2}$.]{\includegraphics[width=0.35\linewidth]{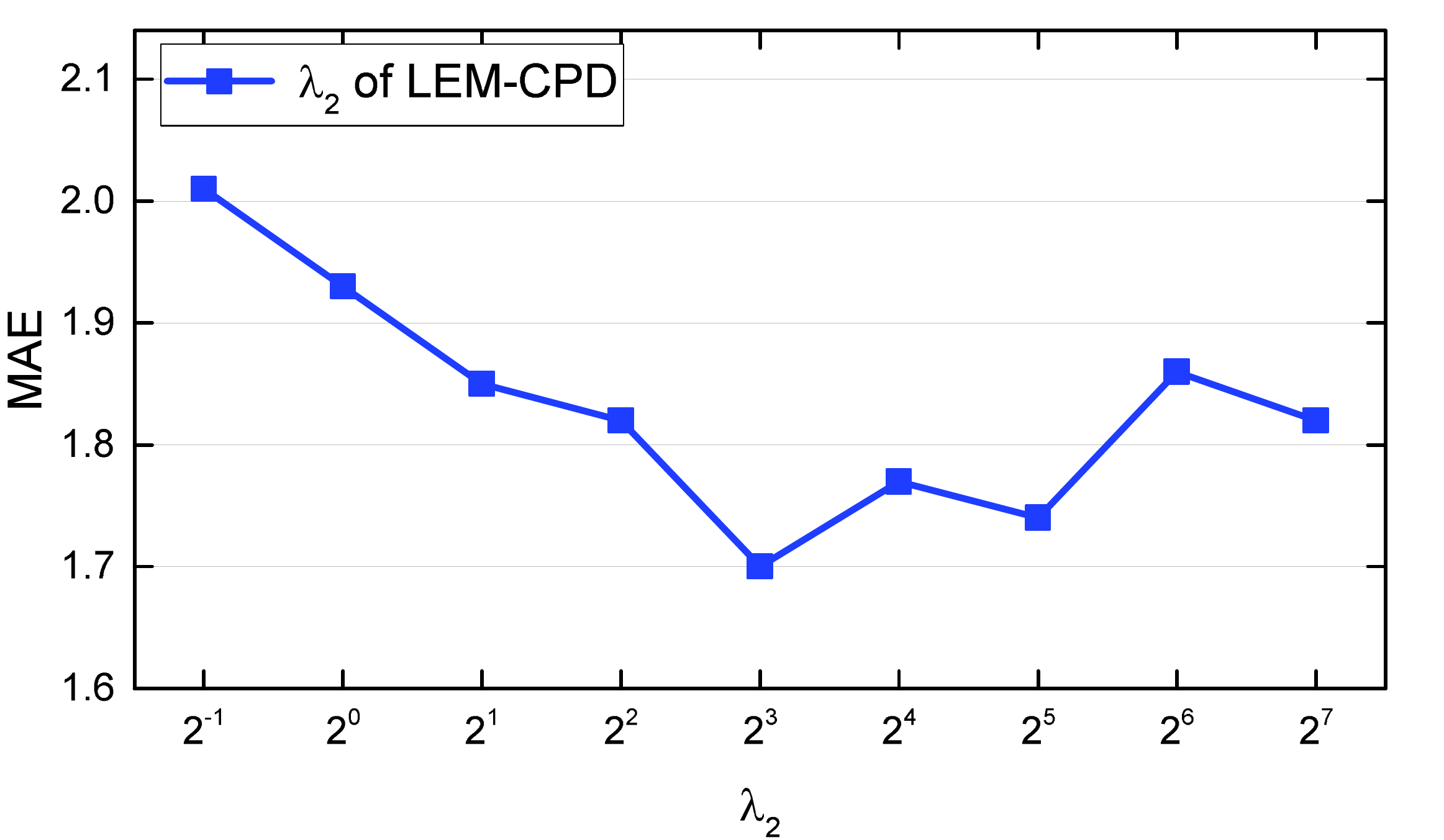}} 
		\caption{Effect of Parameters.} 
		\label{fig-parameters}
	\end{figure}

	This section evaluates the performances of LEM-CPD by varying the critical parameters ($\alpha$, $\lambda_{1}$ and $\lambda_{2}$). We here show the experimental results for the Flow dataset. We discuss them separately but pick them up by the grid search method because four parameters have high dimensional correlations that are hard to visualize. Our illustration approach that discusses parameters separately has been widely used in many other research papers \cite{deng2016latent,gong2018network}.

	Fig. \ref{fig-parameters} (a) shows the different performances with a varying setting for $\alpha$. 
	The $\alpha$ indicates the balance factor to control the importance of prediction results. When we increase $\alpha$ from 0.1 to 0.6, the results improve significantly. However, the performance tends to stay stable at 0.5 $\leq$ $\alpha$ $\leq $ 0.9. In particular, LEM-CPD achieves the best result when $\alpha = 0.6$, while it can get good performance if the $\alpha$ is set between 0.5 and 0.9.  It indicates that the prediction result is important and useful to the change point detection.
	
	Fig. \ref{fig-parameters} (b) and (c) reveal the effect of varying $\lambda_{1}$ and $\lambda_{2}$. These two parameters determine the strength of the transition matrices and the long-term guidance, respectively. $\lambda_{1}$=$2^{-1}$ and $\lambda_{2}$=$2^{3}$ yield the best prediction results for LEM-CPD.

	\section{Conclusion}
	To date, despite achievements in graph-based CPD, most recent methods cannot perform well in the real-world dynamic networks. It is mainly because the real temporal network usually evolves over time. In this paper, we first cast the CPD problem as a prediction task, and develop a novel method to capture latent transitions from previous timestamps to the next. When we made a prediction for the target network, the change point can be identified by a trade-off strategy. Compared with other approaches, the main innovation of this paper is providing an effectiveness CPD method via a prediction process. All the experimental results demonstrate the superiority of LEM-CPD, which achieves more 12 percent than other competitive methods.

	\bibliography{bibliography}

\begin{thebibliography}{10}
\expandafter\ifx\csname url\endcsname\relax
  \def\url#1{\texttt{#1}}\fi
\expandafter\ifx\csname urlprefix\endcsname\relax\def\urlprefix{URL }\fi
\expandafter\ifx\csname href\endcsname\relax
  \def\href#1#2{#2} \def\path#1{#1}\fi

\bibitem{wang2020dynamic}
X.~Wang, Y.~Lu, C.~Shi, R.~Wang, P.~Cui, S.~Mou, Dynamic heterogeneous
  information network embedding with meta-path based proximity, IEEE
  Transactions on Knowledge and Data Engineering.

\bibitem{chen2019periodicity}
J.~Chen, K.~Li, H.~Rong, K.~Bilal, K.~Li, S.~Y. Philip, A periodicity-based
  parallel time series prediction algorithm in cloud computing environments,
  Information Sciences 496 (2019) 506--537.

\bibitem{wang2017fast}
Y.~Wang, A.~Chakrabarti, D.~Sivakoff, S.~Parthasarathy, Fast change point
  detection on dynamic social networks., in: Proceedings of 26th International
  Joint Conference on Artificial Intelligence, 2017, pp. 2992--2998.

\bibitem{gong2018network}
Y.~Gong, Z.~Li, J.~Zhang, W.~Liu, Y.~Zheng, C.~Kirsch, Network-wide crowd flow
  prediction of sydney trains via customized online non-negative matrix
  factorization, in: Proceedings of the 27th ACM International Conference on
  Information and Knowledge Management, 2018, pp. 1243--1252.

\bibitem{wang2019deep}
B.~Wang, J.~Lu, Z.~Yan, H.~Luo, T.~Li, Y.~Zheng, G.~Zhang, Deep uncertainty
  quantification: A machine learning approach for weather forecasting, in:
  Proceedings of the 25th ACM SIGKDD International Conference on Knowledge
  Discovery \& Data Mining, 2019, pp. 2087--2095.

\bibitem{chen2021dynamic}
J.~Chen, K.~Li, K.~Li, P.~S. Yu, Z.~Zeng, Dynamic planning of bicycle stations
  in dockless public bicycle-sharing system using gated graph neural network,
  ACM Transactions on Intelligent Systems and Technology (TIST) 12~(2) (2021)
  1--22.

\bibitem{huang2020laplacian}
S.~Huang, Y.~Hitti, G.~Rabusseau, R.~Rabbany, Laplacian change point detection
  for dynamic graphs, in: Proceedings of the 26th ACM SIGKDD, 2020, pp.
  349--358.

\bibitem{huang2021scalable}
S.~Huang, G.~Rabusseau, R.~Rabbany, Scalable change point detection for dynamic
  graphs (2021) 1--6.

\bibitem{ide2004eigenspace}
T.~Id{\'e}, H.~Kashima, Eigenspace-based anomaly detection in computer systems,
  in: Proceedings of the tenth ACM SIGKDD international conference on Knowledge
  discovery and data mining, 2004, pp. 440--449.

\bibitem{koutra2012tensorsplat}
D.~Koutra, E.~E. Papalexakis, C.~Faloutsos, Tensorsplat: Spotting latent
  anomalies in time, in: 2012 16th Panhellenic Conference on Informatics, IEEE,
  2012, pp. 144--149.

\bibitem{li2014performance}
K.~Li, W.~Yang, K.~Li, Performance analysis and optimization for spmv on gpu
  using probabilistic modeling, IEEE Transactions on Parallel and Distributed
  Systems 26~(1) (2014) 196--205.

\bibitem{chen2020citywide}
C.~Chen, K.~Li, S.~G. Teo, X.~Zou, K.~Li, Z.~Zeng, Citywide traffic flow
  prediction based on multiple gated spatio-temporal convolutional neural
  networks, ACM Transactions on Knowledge Discovery from Data (TKDD) 14~(4)
  (2020) 1--23.

\bibitem{qu2022forecasting}
H.~Qu, Y.~Gong, M.~Chen, J.~Zhang, Y.~Zheng, Y.~Yin, Forecasting fine-grained
  urban flows via spatio-temporal contrastive self-supervision, IEEE
  Transactions on Knowledge and Data Engineering.

\bibitem{liu2022traffic}
J.~Liu, H.~Qu, M.~Chen, Y.~Gong, Traffic flow prediction based on
  spatio-temporal attention block, in: 2022 International Conference on Machine
  Learning, Cloud Computing and Intelligent Mining (MLCCIM), IEEE, 2022, pp.
  32--39.

\bibitem{wang2021optimal}
D.~Wang, Y.~Yu, A.~Rinaldo, Optimal change point detection and localization in
  sparse dynamic networks, The Annals of Statistics 49~(1) (2021) 203--232.

\bibitem{deng2016latent}
D.~Deng, C.~Shahabi, U.~Demiryurek, L.~Zhu, R.~Yu, Y.~Liu, Latent space model
  for road networks to predict time-varying traffic, in: Proceedings of the
  22nd ACM SIGKDD international conference on Knowledge discovery and data
  mining, ACM, 2016, pp. 1525--1534.

\bibitem{gong2020online}
Y.~Gong, Z.~Li, J.~Zhang, W.~Liu, Y.~Zheng, Online spatio-temporal crowd flow
  distribution prediction for complex metro system, IEEE Transactions on
  Knowledge and Data Engineering.

\bibitem{dong2014comparisons}
X.~Dong, Y.~Gong, L.~Zhao, Comparisons of typical algorithms in negative
  sequential pattern mining, in: 2014 IEEE Workshop on Electronics, Computer
  and Applications, IEEE, 2014, pp. 387--390.

\bibitem{gong2015research}
Y.~Gong, C.~Liu, X.~Dong, Research on typical algorithms in negative sequential
  pattern mining, The Open Automation and Control Systems Journal 7~(1).

\bibitem{gong2017nspfi}
Y.~Gong, T.~Xu, X.~Dong, G.~Lv, e-nspfi: Efficient mining negative sequential
  pattern from both frequent and infrequent positive sequential patterns,
  International Journal of Pattern Recognition and Artificial Intelligence
  31~(02) (2017) 1750002.

\bibitem{huang2022series}
J.~Huang, L.~Zhang, Y.~Gong, J.~Zhang, X.~Nie, Y.~Yin, Series photo selection
  via multi-view graph learning, arXiv preprint arXiv:2203.09736.

\bibitem{peel2015detecting}
L.~Peel, A.~Clauset, Detecting change points in the large-scale structure of
  evolving networks, in: Proceedings of the AAAI Conference on Artificial
  Intelligence, Vol.~29, 2015.

\bibitem{zhang2012overlapping}
Y.~Zhang, D.-Y. Yeung, Overlapping community detection via bounded nonnegative
  matrix tri-factorization, in: Proceedings of the 18th ACM SIGKDD
  international conference on Knowledge discovery and data mining, 2012, pp.
  606--614.

\bibitem{xu2014dynamic}
K.~S. Xu, A.~O. Hero, Dynamic stochastic blockmodels for time-evolving social
  networks, IEEE Journal of Selected Topics in Signal Processing 8~(4) (2014)
  552--562.

\bibitem{dong2018f}
X.~Dong, Y.~Gong, L.~Cao, F-nsp+: A fast negative sequential patterns mining
  method with self-adaptive data storage, Pattern Recognition 84 (2018) 13--27.

\bibitem{dong2018rnsp}
X.~Dong, Y.~Gong, L.~Cao, e-rnsp: An efficient method for mining repetition
  negative sequential patterns, IEEE transactions on cybernetics 50~(5) (2018)
  2084--2096.

\bibitem{qiu2021efficient}
P.~Qiu, Y.~Gong, Y.~Zhao, L.~Cao, C.~Zhang, X.~Dong, An efficient method for
  modeling nonoccurring behaviors by negative sequential patterns with loose
  constraints, IEEE Transactions on Neural Networks and Learning Systems.

\bibitem{gong2021inferring}
Y.~Gong, J.~Yi, D.-D. Chen, J.~Zhang, J.~Zhou, Z.~Zhou, Inferring the
  importance of product appearance with semi-supervised multi-modal
  enhancement: A step towards the screenless retailing, in: Proceedings of the
  29th ACM International Conference on Multimedia, 2021, pp. 1120--1128.

\bibitem{li2019sample}
Z.~Li, J.~Zhang, Q.~Wu, Y.~Gong, J.~Yi, C.~Kirsch, Sample adaptive multiple
  kernel learning for failure prediction of railway points, in: Proceedings of
  the 25th ACM SIGKDD International Conference on Knowledge Discovery \& Data
  Mining, 2019, pp. 2848--2856.

\bibitem{li2020field}
Z.~Li, J.~Zhang, Y.~Gong, Y.~Yao, Q.~Wu, Field-wise learning for multi-field
  categorical data, Advances in Neural Information Processing Systems 33 (2020)
  9890--9899.

\bibitem{sarkar2007latent}
P.~Sarkar, S.~M. Siddiqi, G.~J. Gordon, A latent space approach to dynamic
  embedding of co-occurrence data, in: Artificial Intelligence and Statistics,
  PMLR, 2007, pp. 420--427.

\bibitem{thida2013laplacian}
M.~Thida, H.-L. Eng, P.~Remagnino, Laplacian eigenmap with temporal constraints
  for local abnormality detection in crowded scenes, IEEE Transactions on
  Cybernetics 43~(6) (2013) 2147--2156.

\bibitem{jing2018high}
P.~Jing, Y.~Su, X.~Jin, C.~Zhang, High-order temporal correlation model
  learning for time-series prediction, IEEE transactions on cybernetics 49~(6)
  (2018) 2385--2397.

\bibitem{wang2011efficient}
F.~Wang, C.~Tan, P.~Li, A.~C. K{\"o}nig, Efficient document clustering via
  online nonnegative matrix factorizations, in: Proceedings of the 2011 SIAM
  International Conference on Data Mining, SIAM, 2011, pp. 908--919.

\bibitem{gong2020potential}
Y.~Gong, Z.~Li, J.~Zhang, W.~Liu, J.~Yi, Potential passenger flow prediction: A
  novel study for urban transportation development, in: Proceedings of the AAAI
  Conference on Artificial Intelligence, Vol.~34, 2020, pp. 4020--4027.

\bibitem{zhang2015abnormal}
Z.~Zhang, X.~Mei, B.~Xiao, Abnormal event detection via compact low-rank sparse
  learning, IEEE Intelligent Systems 31~(2) (2015) 29--36.

\bibitem{pei2015nonnegative}
Y.~Pei, N.~Chakraborty, K.~Sycara, Nonnegative matrix tri-factorization with
  graph regularization for community detection in social networks, in:
  Twenty-fourth international joint conference on artificial intelligence,
  2015, pp. 2083--2089.

\bibitem{li2009non}
T.~Li, Y.~Zhang, V.~Sindhwani, A non-negative matrix tri-factorization approach
  to sentiment classification with lexical prior knowledge, in: Proceedings of
  the Joint Conference of the 47th Annual Meeting of the ACL and the 4th
  International Joint Conference on Natural Language Processing of the AFNLP,
  2009, pp. 244--252.

\bibitem{wang2011fast}
H.~Wang, F.~Nie, H.~Huang, F.~Makedon, Fast nonnegative matrix
  tri-factorization for large-scale data co-clustering, in: Twenty-Second
  International Joint Conference on Artificial Intelligence, 2011, pp.
  1553--1558.

\bibitem{li2016graph}
X.~Li, G.~Cui, Y.~Dong, Graph regularized non-negative low-rank matrix
  factorization for image clustering, IEEE transactions on cybernetics 47~(11)
  (2016) 3840--3853.

\bibitem{luo2019temporal}
X.~Luo, H.~Wu, H.~Yuan, M.~Zhou, Temporal pattern-aware qos prediction via
  biased non-negative latent factorization of tensors, IEEE transactions on
  cybernetics 50~(5) (2019) 1798--1809.

\bibitem{zhao2017multi}
J.~Zhao, X.~Xie, X.~Xu, S.~Sun, Multi-view learning overview: Recent progress
  and new challenges, Information Fusion 38 (2017) 43--54.

\bibitem{gongspatial}
Y.~Gong, Z.~Li, J.~Zhang, W.~Liu, B.~Chen, X.~Dong, A spatial missing value
  imputation method for multi-view urban statistical data., in: Proceedings of
  29th International Joint Conference on Artificial Intelligence, 2020, pp.
  1310--1316.

\bibitem{gong2021missing}
Y.~Gong, Z.~Li, J.~Zhang, W.~Liu, Y.~Yin, Y.~Zheng, Missing value imputation
  for multi-view urban statistical data via spatial correlation learning, IEEE
  Transactions on Knowledge and Data Engineering.

\bibitem{lee2001algorithms}
D.~D. Lee, H.~S. Seung, Algorithms for non-negative matrix factorization, in:
  Advances in neural information processing systems, 2001, pp. 556--562.

\bibitem{wilson2008study}
R.~C. Wilson, P.~Zhu, A study of graph spectra for comparing graphs and trees,
  Pattern Recognition 41~(9) (2008) 2833--2841.

\bibitem{chung2005laplacians}
F.~Chung, Laplacians and the cheeger inequality for directed graphs, Annals of
  Combinatorics 9~(1) (2005) 1--19.

\bibitem{pillai2005perron}
S.~U. Pillai, T.~Suel, S.~Cha, The perron-frobenius theorem: some of its
  applications, IEEE Signal Processing Magazine 22~(2) (2005) 62--75.

\bibitem{holland1983stochastic}
P.~W. Holland, K.~B. Laskey, S.~Leinhardt, Stochastic blockmodels: First steps,
  Social networks 5~(2) (1983) 109--137.

\bibitem{fowler2006legislative}
J.~H. Fowler, Legislative cosponsorship networks in the us house and senate,
  Social Networks 28~(4) (2006) 454--465.

\bibitem{rasmussen2003gaussian}
C.~E. Rasmussen, Gaussian processes in machine learning, in: Summer School on
  Machine Learning, Springer, 2003, pp. 63--71.

\bibitem{gong2020network}
Y.~Gong, Network-wide spatio-temporal predictive learning for the intelligent
  transportation system, Ph.D. thesis (2020).

\bibitem{petersen2008matrix}
K.~B. Petersen, M.~S. Pedersen, et~al., The matrix cookbook, Technical
  University of Denmark 7~(15) (2008) 510.

\bibitem{gupta2011non}
M.~D. Gupta, J.~Xiao, Non-negative matrix factorization as a feature selection
  tool for maximum margin classifiers, in: CVPR 2011, IEEE, 2011, pp.
  2841--2848.

\end{thebibliography}
	\bibliographystyle{elsarticle-num}
	
	\appendix
	
	\section{Derivation process of Update Rules}
	\label{derivation}
	
	We here derive the update rule of $U_t$ as an example, other variables can be solved with a similar process. The objective of $\mathcal{L}$ could be rewritten as follows:
	
	$\mathcal{L} = L_0 + L_1 + L_2$
	, where: 
	
	\begin{equation}
		\begin{split}
			&L_0 = \sum_{t=1}^{T} ||\bm G_t - \bm U_t\bm C\bm V_t)||_{F}^{2},
			\\& L_1 = \lambda_1 \sum_{t=2}^{T}(||\bm U_{t} - \bm U_{t-1}\bm A ||_{F}^{2}+
			||\bm V_{t} -\bm V_{t-1}\bm B ||_{F}^{2}),\\& L_2 = \lambda_2 (||\bm U^{lt} - \bm U_{T}\bm A||_{F}^{2}+||\bm H^{lt} -\bm H_{T}\bm B ||_{F}^{2}).
		\end{split}
	\end{equation}
	
	We provide the derivative of $L_0$ respect to $\bm U_t$ as an example, the other components can be derived in the same way. $L_0$ could also be rewritten as follows:
	
	\begin{equation}
		\small
		L_0 = \sum_{t=1}^{T} tr( (\bm G_t - \bm U_t\bm C\bm V_t)(\bm G_t - \bm U_t\bm C\bm V_t)^T ),
	\end{equation}
	analogously, we can get:
	
	\begin{equation}
		\small
		L_1 = \lambda_1 \sum_{t=2}^{T}( tr((\bm U_{t} - \bm U_{t-1}\bm A)(\bm U_{t} - \bm U_{t-1}\bm A)^T) +
		||\bm V_{t} -\bm V_{t-1}\bm B ||_{F}^{2}),
	\end{equation}
	
	\begin{equation}
		\small
		L_2 = \lambda_2 ( tr((\bm U^{lt} - \bm U_{T}\bm A)(\bm U^{lt} - \bm U_{T}A)^T) +
		||\bm V^{lt} -\bm V_{T}\bm B ||_{F}^{2}).
	\end{equation}
	
	Taking the derivation of $\mathcal{L}$ with respect to $U_t$, we can get $g(U_t)$ according to \cite{petersen2008matrix}:
	
	As introduced in \cite{lee2001algorithms}, the traditional gradient descent method is expressed as: $U_t$ = $U_t$ - $\gamma g(U_t)$ = $U_t$ - $\gamma (P_{item} + N_{item})$, where $P_{item}$ and $N_{item}$ denote all positive and negative items in $g(U_t)$, respectively (e.g., $P_{item}$ = $G_{t}(CV_t)^{\top}+\lambda _{1}(U_{t-1}A+U_{t+1}A^{\top})+\hat{\lambda _{2}}U^{lt}A^{\top}$). We can set the step $\gamma$ to:
	
	\begin{small}
		\begin{equation}
			\gamma = \frac{U_t}{P_{item}},
		\end{equation}
	\end{small}
	then, we got the update rule of $U_t$ as shown in Equation (8) of the main paper.

	\section{Proof of Convergence}
	\label{convergence}
	
	To prove \textbf{Theorem 1}, we will find an auxiliary
	function similar to that used in the  \cite{lee2001algorithms}. We here give the
	convergence proof of $V_t$ and other variables can similarly be proofed.

	\textbf{Definition 1.} \textit{$G(v,v^{'})$ is an auxiliary function for our final function $\mathcal{L}(v)$ if the following conditions are satisfied:}
	
	\begin{equation}
		G\left(v^{\prime}, v\right) \geq \mathcal{L}(v) \quad \text { and } \quad G(v, v)=\mathcal{J}(v).
	\end{equation}

	\textbf{Lemma 1} \textit{If $G$ is an auxiliary function, then $\mathcal{J}$ is non-increasing under the update:}
	
	\begin{equation}
		\label{eq_G}
		v^{t+1}=\arg \min _{v} G\left(v, v^{t}\right),
	\end{equation}
	
	\textit{consequently, we have:}
	
	\begin{equation}
		\mathcal{L}\left(v^{t+1}\right) \leq G\left(v^{t+1}, v^{t}\right) \leq G\left(v^{t}, v^{t}\right)=\mathcal{L}\left(v^{v}\right).
	\end{equation}
	
	The proof of Lemma 1 is given by \cite{lee2001algorithms}. Lemma 1 illustrates that $\mathcal{L}\left(v^{t+1}\right) \leq \mathcal{L}(v^t)$ when exits $G\left(v, v^{t}\right)$; $v^t$ is a column vector of $V_t$.

	\textbf{Lemma 2.} \textit{If $K(v^{t})$ is a diagonal matrix under the following definition,}
	
	\begin{equation}
		K(v^{t}) = diag((UC)(UC)^Tv./v),
	\end{equation}
	\textit{ then, }
	
	\begin{equation}
		\label{eq_5}
		\begin{split}
			G\left(v, v^{t}\right)=&\mathcal{L}\left(v^{t}\right)+\left(v-v^{t}\right)^{T} \nabla \mathcal{L}\left(v^{t}\right)\\&+\frac{1}{2}\left(v-v^{t}\right)^{T} K\left(v^{t}\right)\left(v-v^{t}\right),
		\end{split}
	\end{equation}
	\textit{is an auxiliary function for $\mathcal{L}(v)$.}
	
	\begin{proof}
		Since $G(v,v)$ = $\mathcal{L}(v)$ is obvious, we need only show that $G(v,v^{t}) \geq  \mathcal{L}(v)$. To do this, we compare 
		
		\begin{equation}
			\begin{split}
				\mathcal{L}(v)=&\mathcal{L}\left(v^{t}\right)+\left(v-v^{t}\right)^{T} \nabla \mathcal{L}\left(v^{t}\right)\\&+\frac{1}{2}\left(v-v^{t}\right)^{T}\left((UC)(UC)^T\right)\left(v-v^{t}\right),
			\end{split}
		\end{equation}
		with Equation (\ref{eq_5}) to find that $G(v,v^{t}) \geq  \mathcal{L}(v)$ is equivalent to
		
		\begin{equation}
			0 \leq\left(v-v^{t}\right)^{T}\left[K\left(v^{t}\right)-(UC)(UC)^T\right]\left(v-v^{t}\right).
		\end{equation}

		The next step is to prove $\left[K\left(v^{t}\right)-(UC)(UC)^{T}\right]$ is positive semi-definite. Let $Q = (UC)(UC)^{T}$, then $\left[K\left(v^{t}\right)-(UC)(UC)^{T} \right]$ can be expressed as $[diag(Qh./h) - Q]$. As the \textbf{Lemma 1} provided in \cite{gupta2011non}, if Q is a symmetric non-negative matrix and $v$ be a positive vector, then the matrix $\hat{Q} = diag(Qv./v) - Q \succeq 0$.

		Replacing $G(v,v^{t})$ in Equation (\ref{eq_G}) by Equation (\ref{eq_5}) results in the update rule:
		
		\begin{equation}
			v^{t+1}=v^{t}-K\left(v^{t}\right)^{-1} \nabla \mathcal{L}\left(v^{t}\right).
		\end{equation}
		
		Since Equation (\ref{eq_5}) is an auxiliary function, $\mathcal{L}$ is nonincreasing under this update rule, according
		to \textbf{Lemma 1}. Writing the components of this equation explicitly, we obtain

		\begin{equation}
			v_{a}^{t+1}=v_{a}^{t} \frac{\left(UC g\right)_{a}}{\left(UC ( (UC)^{T} v\right))_{a}}.
		\end{equation}

		By reversing the roles of $U$ and $V$ in \textbf{Lemma 1} and \textbf{Lemma 2}, $\mathcal{L}$ can similarly be shown to be nonincreasing under the update rules for $U$.
		
	\end{proof}

	\begin{figure}[t]
		\centering
		\includegraphics[width=0.5\linewidth]{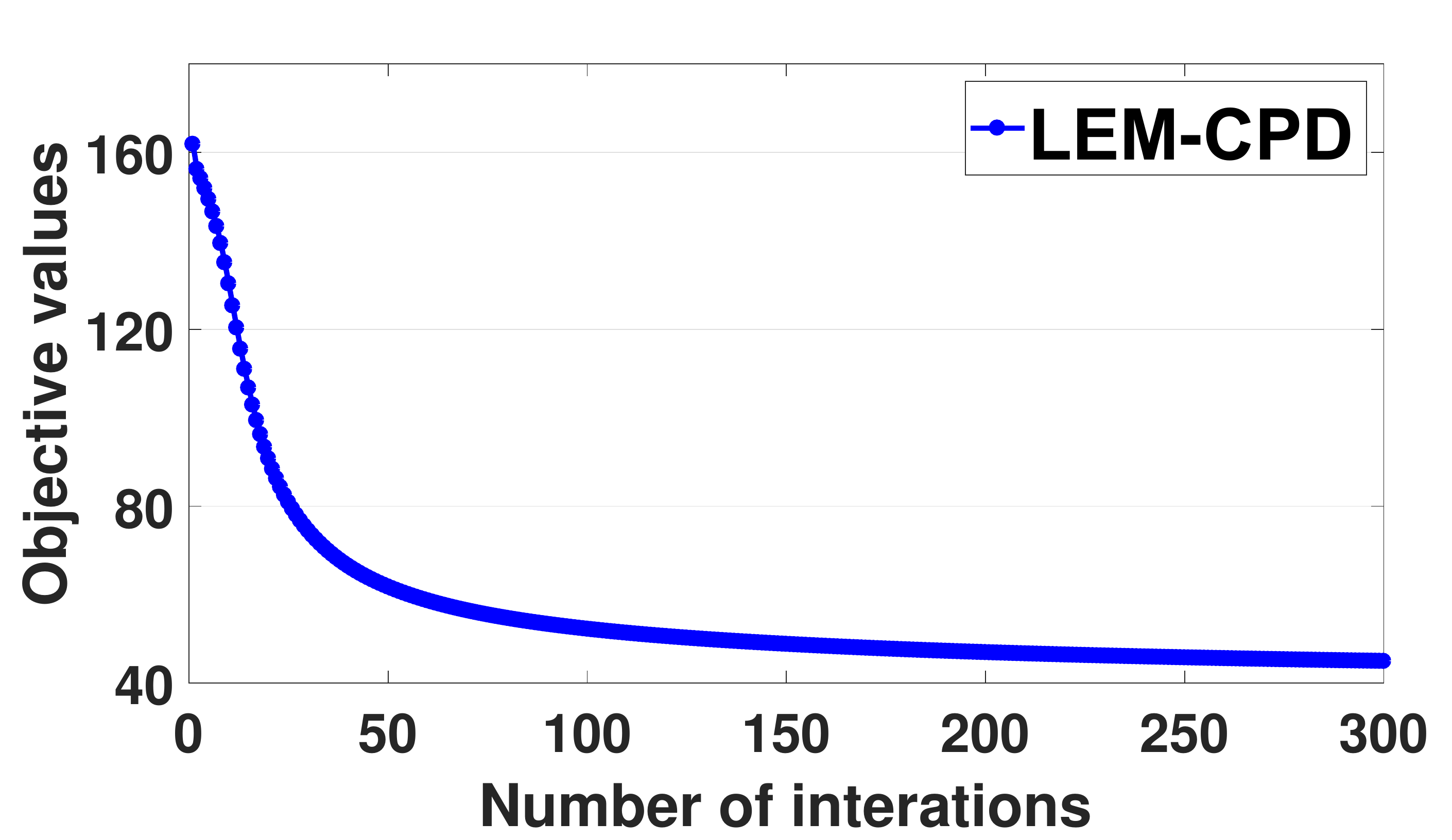}
		\caption{Convergence rate in the Flow dataset.}	
		\label{figure-convergence}
	\end{figure}
	
	As shown in Figures \ref{figure-convergence}, our model LEM-CPD can efficiently converge into a local optimization with a small number of iterations.

\end{document}